# A FORCE-FREE MAGNETIC FIELD SOLUTION IN TOROIDAL COORDINATES

**Gerald E. Marsh**
Argonne National Laboratory (Ret)
gemarsh@uchicago.edu
Chicago, Illinois, 60615, USA

**ABSTRACT**

While there are no known analytic solutions for force-free fields in toroidal coordinates, with a reasonable boundary condition it is possible to find a solution for the surface field and, with a restriction on the form of the field, to the interior of the torus as well.

**Introduction**

In topology, a torus is a surface of genus one, meaning it has one hole, and the relevant question here is whether it is possible for the surface of a torus to have on it a non-singular force-free magnetic field, meaning one that does not vanish at any point on the surface. Such a field satisfies the force-free magnetic field equation $\nabla \times \boldsymbol{B} = \alpha \boldsymbol{B}$. Arnold[1] has shown that for force-free magnetic fields the field lines will lie on tori provided the field is non-singular and $\alpha$ is not constant. In addition, a theorem by Hopf tells us that that the torus and the Klein bottle are the only smooth, compact, connected surfaces without boundary allowing a vector field without a singularity.[2]

It is worth stating the Poincaré-Hopf theorem somewhat more formally: If a smooth, compact, connected surface $S$ has on it a vector field with only isolated zeros, then its Euler characteristic $\chi(S)$ is an appropriate sum of the index of each zero. Any closed orientable surface is topologically equivalent to a sphere with $p$-handles and Euler characteristic $\chi(S) = 2 - 2p$.

What the Poincaré-Hopf theorem states is that only surfaces with Euler Characteristic zero can have a vector field which is nowhere zero. Only the torus and Klein bottle have Euler characteristic zero. Since real Klein bottles in 3-dimensional space cannot exist, only the torus is relevant.

Below it will be shown that there is a non-singular force-free magnetic field restricted to the surface of a torus, and under a restriction on the form of the field, to the interior as well.

With regard to force-free magnetic fields in plasma physics, where the condition for plasma equilibrium is given by $(\nabla \times \boldsymbol{B}) \times \boldsymbol{B} = \nabla p$, where $p$ is the plasma pressure, the magnetic field will be force-free if $\nabla p = 0$. Force-free means that the "self-force" or Lorentz force vanishes. Force-free magnetic field configurations are difficult to find because $(\nabla \times \boldsymbol{B}) \times \boldsymbol{B} = 0$ is a nonlinear equation. The plasma $\beta$ is defined as the ratio of the plasma pressure to the magnetic pressure $p_m$. The force-free approximation is valid for "low-beta" plasmas.

## Toroidal Coordinates and the Force-Free Relations

Solving for an exact solution to the force-free magnetic field equations in toroidal coordinates is a difficult problem. An extensive history and the approaches used to solve both the exterior and interior toroidal problem has been given by Marsh.[3] In particular, for the interior problem no exact solution is known and one obtains a first order differential equation for $\alpha$, which can most likely only be dealt with by numerical methods.

There are numerous definitions for toroidal coordinates, and the one used here is shown in Fig. 1

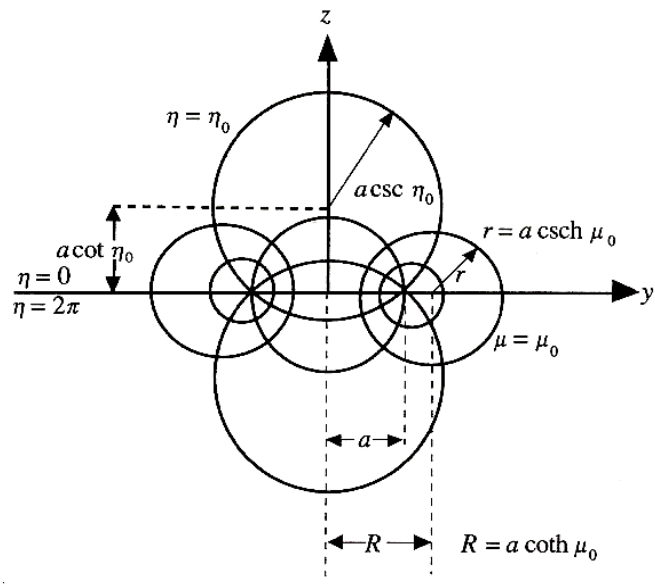

Figure 1. Toroidal coordinates. Note that $a^2 = R^2 - r^2$.

The relation between rectangular coordinates and toroidal coordinates is given by

$$x = \frac{a \sinh\mu \cos\phi}{\cosh\mu - \cos\eta}, \quad y = \frac{a \sinh\mu \sin\phi}{\cosh\mu - \cos\eta}, \quad z = \frac{a \sin\eta}{\cosh\mu - \cos\eta}.$$

(1)

The metric coefficients for the coordinates are then

$$h_\mu = h_\eta = \frac{a}{\cosh\mu - \cos\eta}, \quad h_\phi = \frac{a \sinh\mu}{\cosh\mu - \cos\eta}.$$

(2)

In toroidal coordinates the force-free magnetic field equation, $\nabla \times \boldsymbol{B} = \alpha \boldsymbol{B}$, yields the following three equations

$$h_\eta B_\eta = -\frac{1}{\alpha h_\phi} \partial_\mu (h_\phi B_\phi),$$

$$h_\mu B_\mu = \frac{1}{\alpha h_\phi} \partial_\eta (h_\phi B_\phi),$$

$$\partial_\mu (h_\eta B_\eta) - \partial_\eta (h_\mu B_\mu) = \frac{\alpha h_\eta}{\sinh\mu} (h_\phi B_\phi).$$

(3)

These equations are very general and are applicable to all force-free fields in toroidal coordinates.

The divergence of **B** is given by

$$\nabla \cdot \boldsymbol{B} = \frac{1}{h_\mu h_\eta h_\phi} [\partial_\mu (h_\eta h_\phi B_\mu) + \partial_\eta (h_\phi h_\mu B_\eta) + \partial_\phi (h_\mu h_\eta B_\phi)].$$

(4)

Imposing axial symmetry ($\partial_\phi B_\phi = 0$) and the requirement that $\nabla \cdot \boldsymbol{B} = 0$ results in

$$\partial_\mu (h_\eta h_\phi B_\mu) + \partial_\eta (h_\phi h_\mu B_\eta) = 0.$$

(5)

This means that $\alpha$ is only a function of $h_\phi B_\phi$; that is, $\alpha = \alpha(h_\phi B_\phi)$. The force-free relation also implies that $\nabla \alpha \cdot \boldsymbol{B} = 0$. Since $\alpha$ is not a function of $\phi$ by symmetry, this in turn implies that

$$\partial_\eta \, \alpha = -\frac{B_\mu}{B_\eta} \partial_\mu \alpha.$$

(6)

Combining Eq. (6) with Eqs. (3) yields the differential equation,

$$\partial_\mu \left(\frac{1}{h_\phi} \partial_\mu (h_\phi B_\phi)\right) + \partial_\eta \left(\frac{1}{h_\phi} \partial_\eta (h_\phi B_\phi)\right) + \frac{\partial_\mu \alpha}{\alpha h_\phi} \left(\frac{B_\mu}{B_\eta} \partial_\eta (h_\phi B_\phi) - \partial_\mu (h_\mu B_\mu)\right)$$
$$+ \frac{\alpha^2 h_\eta}{\sinh \mu} (h_\phi B_\phi) = 0.$$

(7)

This equation leads to an intractable equation for $\alpha$, but will be simplified by imposing an additional restriction on $B_\mu$ as discussed below.

## Boundary Conditions

The cylindrically symmetric Lundquist solution to the force-free field equations is shown in Fig. 2. The Lundquist solution[4] is obtained by restricting $\alpha$ to a constant and further restricting the magnetic field to the form $\boldsymbol{B} = [0, B_\phi(r), B_z(r)]$.

The field equations will then give the solution $\boldsymbol{B} = A_0[0, J_1(\alpha r), J_0(\alpha r)]$, where $J_0$ and $J_1$ are Bessel functions and $A_0$ is a constant.

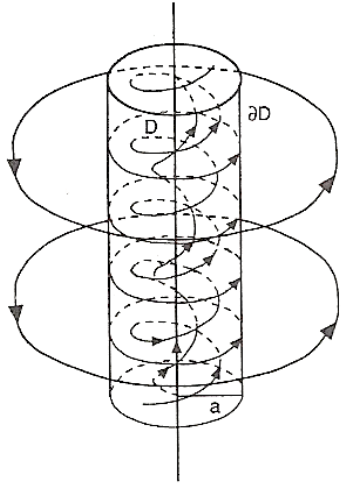

Figure 2. The Lundquist solution. The figure is drawn so that $B_z = J_0(\alpha a) = 0$ on the cylinder $r = a$.

If one chooses to apply the solution $\boldsymbol{B} = A_0[0, J_1(\alpha r), J_0(\alpha r)]$ in a cylindrical region D bounded by $\partial D$ (as shown if Fig.2) such that $J_0(\alpha a) = 0$, the solution matches smoothly to an external field given by $\boldsymbol{B} = [0, (aA_0/r) J_1(\alpha a), 0]$ and no surface currents are required to satisfy the boundary condition.

### The Equation for $\alpha$ on the Surface of the Torus,

The following differential equation for $\alpha$ follows from that of Eq. (7) with the additional requirement that $B_\mu$ vanishes everywhere.

$$\partial_\mu \left( \frac{1}{h_\phi} \partial_\mu (h_\phi B_\phi) \right) - \frac{\partial_\mu \alpha}{\alpha h_\phi} \partial_\mu (h_\phi B_\phi) + \frac{\alpha^2 h_\eta}{\sinh \mu} (h_\phi B_\phi) = 0$$

(8)

If one computes the first two terms of Eq. (8) and adds the third term, it can be seen that $B_\phi$, which is not an explicit function of $\mu$, drops out.

After solving Eq. (8) for $\alpha$ the magnetic field itself will be found by imposing the boundary condition $B_\mu(\mu_0) = 0$.

Equation (8) has two solutions[5]

$$\alpha = \pm\frac{1}{a}[\sqrt{(-\cos^2\eta + \cos\eta\,\cosh\mu + \cos^2\eta\,\coth^2\mu - 2\cos\eta\,\cosh\mu\,\coth^2\mu}$$
$$+ \cosh^2\mu\,\coth^2\mu - \sinh^2\mu)].$$

(9)

When $\eta$ is held constant, say in the positive equation, and $\alpha$ plotted as a function of $\mu$, $\alpha$ grows monotonically with increasing $\mu$. In what follows, it will be seen that the field winds around the torus specified by a particular value of $\mu$ and the sign of $\alpha$ determines the handedness of the field while the period of the twisting field is given by $|\alpha|$.

With reference to Fig. 1, henceforth $\mu = 1$ and $a = 2$ will generally be used. The plot of both solutions given by Eq. (9) for $\alpha$ is shown in Fig. 3.

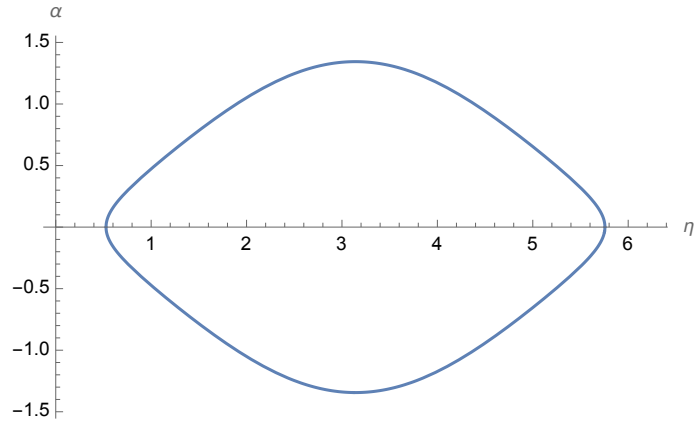

Figure 3. The plot of both solutions for $\alpha$ in Eq. (9) as a function of $\eta$ where $0 \leq \eta \leq 2\pi$.

Figure 3 shows that the solutions for $\alpha$ do not cover the full range of $\eta$ from $0 \leq \eta \leq 2\pi$. For $0 \leq \eta \leq 0.529$ and $5.753 \leq \eta \leq 2\pi$, $\alpha$ is pure imaginary so that the solutions given in Eq. (9) are not applicable. In these regions, the real part of $\alpha$ vanishes and since $\alpha$ must be a real function, the force-free relation $\nabla \times \boldsymbol{B} = \alpha\boldsymbol{B}$ implies that the field $\boldsymbol{B}$ is given by the gradient of a scalar function. It will be seen below that the transition from a force-free field to this gradient field is smooth with **B** always greater than zero so that the field is non-singular.

## The Force-Free Field on the surface of the Torus

Solutions to the force-free field equations may now be found for any torus satisfying the condition that $\mu$ be constant on it. The second of Eqs. (3) with $B_\mu(\mu_0) = 0$ tells us that $h_\phi B_\eta$ is not a function of $\eta$. It is also not a function of $\phi$ by axial symmetry and therefore is only a function of $\mu$, and because $\mu$ is constant on the surface of the torus, $h_\phi B_\phi$ is also constant. Therefore, $B_\phi = C_1/h_\phi$.

In Eq. (4), $\nabla \cdot \boldsymbol{B} = 0$ and cylindrical symmetry along with $B_\mu = 0$ imply that $\partial_\eta(h_\mu h_\phi B_\eta) = 0$ so that $B_\eta = C_2/h_\phi h_\mu$. Figure 4 shows $B_\phi$ and $B_\eta$ for $C1 = 1, C2 = 2$, with $\mu$ =1 and *a*=2.

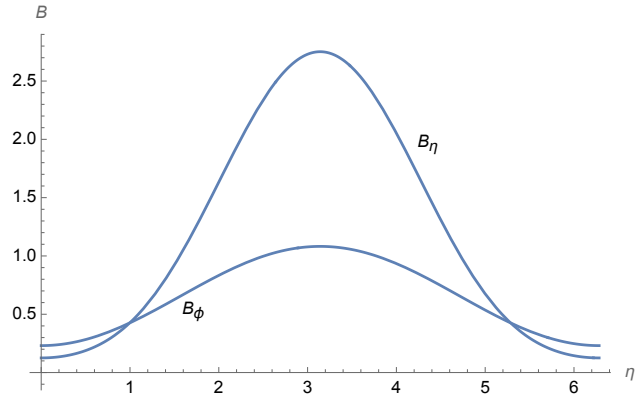

Figure 4. $B_\phi$ and $B_\eta$ plotted for $0 \leq \eta \leq 2\pi$.

Where the curves for $B_\phi$ and $B_\eta$ cross the magnitude of these components are equal so that the angle of their vector is at $\pi/4$ radians with respect to $\hat{\phi}$. The magnitude of the components depends on the choice of the constants C1 and C2.

Note that in the regions $0 \leq \eta \leq 0.529$ and $5.753 \leq \eta \leq 2\pi$, where $\alpha$ is pure imaginary there is no discontinuity in the field components and that the components $B_\phi$ and $B_\eta$ never vanish so that the vector field they represent is not singular. Coupled with the fact that $\alpha$ is not constant, the field satisfies Arnold's requirements for a force-free field on a torus.

The fact that this solution has a smooth transition from a force-free magnetic field to a field given by the gradient of a scalar function in the regions $0 \leq \eta \leq 0.529$ and $5.753 \leq \eta \leq 2\pi$ is one of the most interesting features of the solution.

A vector plot of the vector $\boldsymbol{B} = (0, B_\phi, B_\eta)$ gives a better idea what the field looks like. This is shown in Fig. 5.

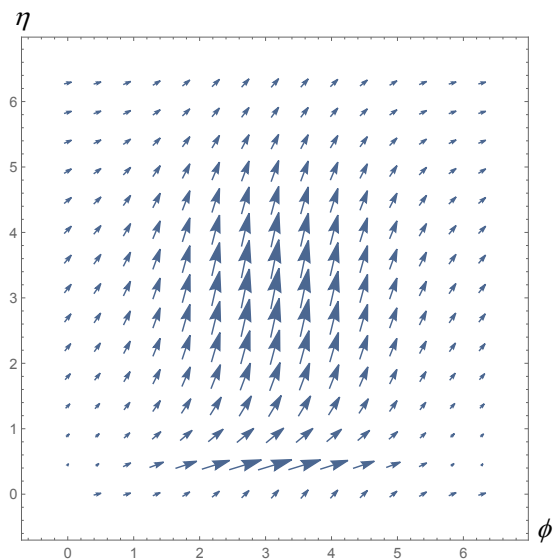

Figure 5. This is a plot of the vector field $\boldsymbol{B} = (0, B_\phi, B_\eta)$ as a function of $\phi$ and $\eta$. The magnitude of the field is given by the length of the arrows. For a given $\eta$ the projection of the vectors on the $\phi$-axis (the $B_\phi$ component), remains constant so that axial symmetry is preserved. The angle of the vectors along a given $\eta$ with the $\phi$-axis changes with $\phi$, although that is somewhat difficult to see in the figure.

This field is unusual since both the pitch and magnitude change with location on the surface of the torus. This should be compared to the Lundquist solution shown in Fig. 2 and its surface at $r = a$.

Considering only the plot of Fig. 5 itself, without the "padding" around it, one can get idea of how the field looks on a torus by identifying the $\phi$ sides of this plot and then identifying the ends of the resulting cylinder.

Alternatively, one can use a stream plot, which loses the magnitude information. The stream plot itself is shown in Fig 6 and its mapping onto the torus in Fig. 7.

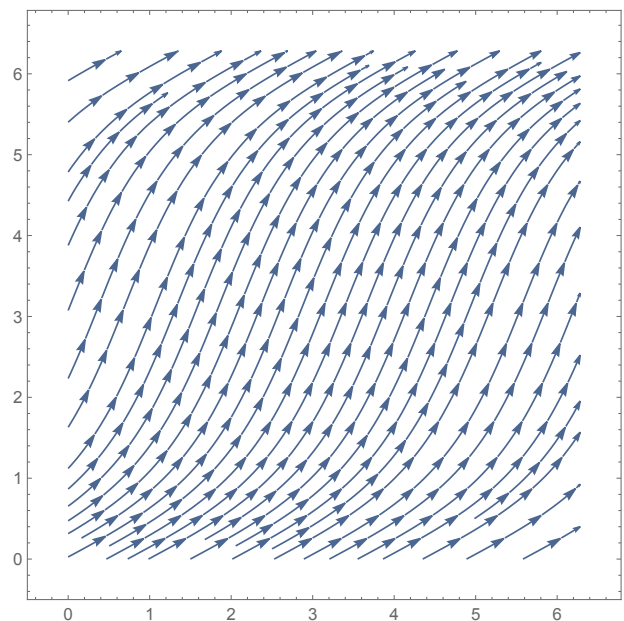

Figure 6. A stream plot of the vector field shown in Fig. 5. The magnitude information of Fig. 5 cannot be made a part of this plot.

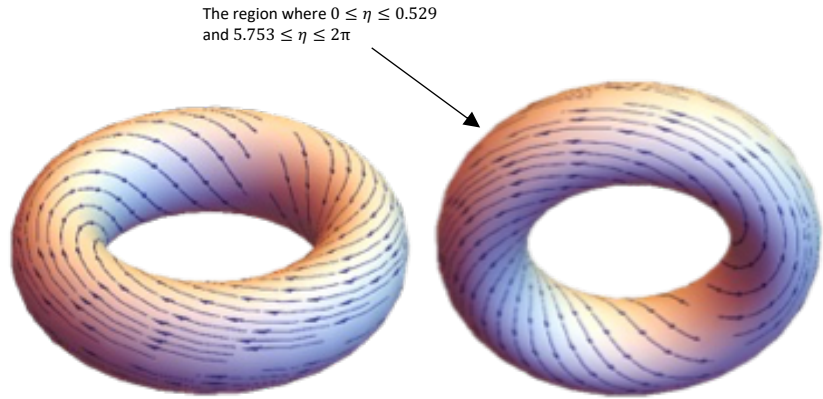

The region where $0 \leq \eta \leq 0.529$ and $5.753 \leq \eta \leq 2\pi$

Figure 7. Two views of the stream plot of Fig. 6 mapped onto the torus. The gap seen in the first figure is an artifact of the mapping and not a discontinuity in the field. The region on the perimeter of the torus where the field becomes a gradient field is also indicated.

In producing the plots in Fig. 7 the axes and "padding" around the stream plot in Fig. 6 were remove before doing the mapping. Unfortunately, the mapping program only recognizes the removal of the axes--hence the gap seen particularly in the first figure. It is not real and only an artifact of the mapping.

## Summary

While the general solution to the force-free field equations in toroidal coordinates remains unknown, the solution for the magnetic field on the surface of a torus when a reasonable, and possibly necessary, boundary condition is assumed has been derived here. It satisfies Arnold's requirement that the force-free magnetic field lines will only lie on tori if the field is non-singular and $\alpha$ is not constant—when $\alpha$ is a constant, force-free fields can have a much more complicated topology (see Ref 3, p.62). The analytic solution found is very interesting because it also has a region where the field becomes the gradient of a scalar function. If, in addition, the form of the field is restricted to $\boldsymbol{B} = [0, B_\phi(r), B_\eta(r)]$ then the solution found applies to the interior of the torus as well.